\documentclass{mn2e}

\input epsf

\usepackage{graphicx}
\usepackage{txfonts}
\usepackage{epsfig}

\title[Asteroseismology of Arcturus]{Asteroseismology of red giants:
photometric observations of Arcturus by \emph{SMEI}}

\author[N.~J.~Tarrant et al.]{N.~J.~Tarrant, W.~J.~Chaplin,
Y.~Elsworth, S.~A.~Spreckley, I.~R.~Stevens\\ School of Physics and
Astronomy, University of Birmingham, Edgbaston, Birmingham B15 2TT,
U.K.}

\begin{document}

\maketitle

\begin{abstract}

We present new results on oscillations of the K1.5\,III giant Arcturus
($\alpha$\,Boo), from analysis of just over 2.5\,yr of precise
photometric observations made by the \emph{Solar Mass Ejection Imager}
(\emph{SMEI}) on board the Coriolis satellite. A strong mode of
oscillation is uncovered by the analysis, having frequency $3.51 \pm
0.03\,\rm \mu Hz$. By fitting its mode peak, we are able offer a
highly constrained direct estimate of the damping time ($\tau = 24 \pm
1\,\rm d$). The data also hint at the possible presence of several
radial-mode overtones, and maybe some non-radial modes. We are also
able to measure the properties of the granulation on the star, with
the characteristic timescale for the granulation estimated to be $\sim
0.50 \pm 0.05\,\rm d$.

\end{abstract}

\begin{keywords}

stars: individual (Arcturus) -- stars: red giants -- stars:
oscillations

\end{keywords}

\section{Introduction}
\label{sec:intro}

Oscillations have now been observed in several G and K-class red giant
stars (see Bedding \& Kjeldsen 2006, and references therein). The
oscillations appear to be Sun-like in nature (Frandsen et al. 2002;
de~Ridder et al. 2006; Barban et al. 2007), meaning they are self
excited stochastically by convection. It had been thought that the
radial acoustic (p) modes would dominate the oscillation spectra
(Christensen-Dalsgaard, 2004).  However, recent observations by Hekker
et al. (2006) suggest non-radial modes may also be present in some
giants.

In this paper we present new results on oscillations of the K1.5\,III
giant Arcturus ($\alpha$\,Boo), from analysis of photometric
observations made by the \emph{Solar Mass Ejection Imager}
(\emph{SMEI}) on board the Coriolis satellite.  Arcturus has been the
subject of several observing campaigns that have studied its
variability. Observations of Arcturus made from the ground in Doppler
velocity (e.g., Innis et al. 1998; Belmonte et al. 1990; Hatzes \&
Cochran 1994; Merline 1995), and from space in photometric intensity
(Retter et al. 2003), have uncovered evidence for variability on
periods of the order of a few days. These periods correspond to
frequencies of a few micro-Hertz, and are close to predictions for
acoustic modes based on estimates of the mass, radius and effective
temperature of Arcturus (see Section~\ref{sec:data} below).

While Arcturus is, indisputably, a variable star, no studies to date
have presented convincing evidence for the identification of
\emph{individual} modes of oscillation.  This can be something of a
non-trivial problem for red giants. First, spacings in frequency
between radial overtones are expected to be small. This is because the
spacings are predicted to scale, to first order, with the square root
of the mean densities of the stars. The predicted spacing for

Arcturus, based on estimates of its mass and radius, is only $\approx
0.9\,\rm \mu Hz$. Such a small spacing places demands on the length of
dataset required to resolve peaks cleanly in frequency. Second, if the
damping of stochastically excited modes is sufficiently strong it can
give peaks widths that may cause neighbouring modes to overlap in
frequency (see Stello et al. 2006). Third, if non-radial modes are
observed at prominent levels they will give rise to additional
crowding of the mode spectrum. Some of these non-radial modes may be
expected to possess mixed acoustic and gravity-mode-like
characteristics. The spacings in frequency of these mixed modes can be
somewhat irregular (Christensen-Dalsgaard, 2004) creating further
complications for mode identification, particularly in short datasets,
if they appear at observable levels.

The \emph{SMEI} observations that we present here extend over a period
lasting just over two-and-a-half years, and have a reasonable overall
duty cycle. The excellent resolution possible in frequency is
unprecedented for near-continuous, high-cadence photometric studies of
red giant variability, and we are able to identify, for the first
time, individual modes of oscillation of Arcturus. We offer a direct
estimate of the mode damping time. And our results also hint at the
possible presence of several radial-mode overtones, and maybe some
non-radial modes. Finally, we are able to measure the granulation
properties of the star.

\section{Data and lightcurve extraction}
\label{sec:data}

The Coriolis satellite lies in a Sun-synchronous polar orbit of period
$\sim 101$\,minutes. \emph{SMEI} comprises 3 cameras, each with a
field of view of $60 \times 3$ degrees. The cameras are aligned such
that the instantaneous total field of view is a strip of sky of size
$170 \times 3$ degrees; a near-complete image of the sky is obtained
from data on all three cameras after every orbit. Individual images --
which are made from 10 stacked exposures, with a total integration
time of about 40\,sec -- occupy an arc-shaped $1242 \times 256$-pixel
section of each $1272 \times 576$-pixel CCD.  Observations are made in
white light, and the spectral response of the cameras is very broad,
extending from $\sim 500$ to $\sim 900\,\rm nm$, with a peak at about
700\,nm. A detailed description of the data analysis pipeline used to
generate the lightcurves may be found in Spreckley \& Stevens (in
preparation). Here, we give a brief summary of the main steps.

Poor-quality frames having high background are first excluded from any
further processing. Processing of the good frames begins with
subtraction of bias, calculated from overscan regions at the edges of
each frame, and a temperature-scaled dark-current signal. The frames
are then flat fielded and spurious signals from cosmic ray hits are
removed from the images. The Camera \#2 data suffer from some stray
light, and further cleaning of these data is performed to minimize the
stray-light contribution (Buffington, private communication). The
stray-light problem is concentrated in a small number of pixels on the
CCD, and is not a major cause of concern for the data collected on
Arcturus. Once the images have been cleaned aperture photometry is
performed with a modified version of the \emph{DAOPHOT} routines
(Stetson 1987). The target star is tracked, and its lightcurve is
corrected for the degradation of the CCD over the course of the
mission, and a position-dependent correction is applied to compensate
for variation of the Point Spread Function (PSF) across the
frames. When the star lies within the field of view of one of the
cameras, a single photometric measurement of its intensity is
therefore obtained once every orbit.

We have obtained lightcurves of Arcturus from the Camera \#1 and
Camera \#2 data. (There are some problems with data from Camera \#3,
which points closest to the Sun.) The Camera \#1 lightcurve, which

begins 2003 April 10, comprises three data segments when Arcturus was
visible, each having a length up to 120 days, with $\approx\,240$-day
data gaps between each segment. Within each segment 50 to 70\,per cent
of data points are usable, with a fill of 25.6\,per cent over the
entire 812.2-day length. The Camera \#2 lightcurve, which begins 2003
June 24, comprises six segments of observations each having a length
around 60 days, separated by $\approx\,120$-day gaps. Apart from the
shorter final segment, 60 to 80\,per cent of the data points are
usable in each segment, with a fill of 23.1\,per cent over the entire
975.7-day length.


 \begin{figure}
 \centerline{\epsfxsize=7.25cm\epsfbox{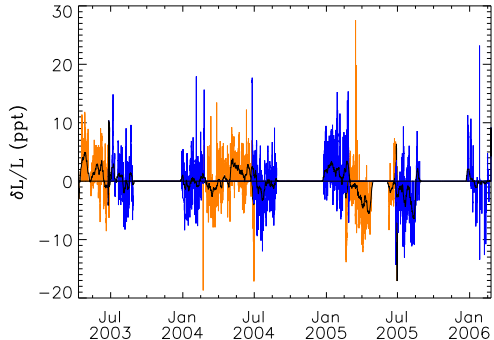}}
 
 \caption{Residual lightcurves, in ppt, of the Camera \#1 (orange) and
  Camera \#2 (blue) data. The dark solid line shows the result of
  smoothing with a boxcar filter of width $\sim 7\,\rm d$.}

 \label{fig:res}
 \end{figure}


Fig.~\ref{fig:res} shows the residual lightcurves of both cameras, in
units of parts per thousand (ppt). The residual lightcurves for each
camera were calculated, independently, as follows. The mean intensity
was first computed for the usable data in all segments of a given
camera. Fractional variations, $\delta L / L$, were then calculated
with respect to this mean intensity. No adjustments were made to match
the residuals from the two cameras.  The Camera \#1 lightcurve
residuals are plotted in orange, and the Camera \#2 residuals are
plotted in blue. The dark solid line shows the result of smoothing
with a boxcar filter of width $\sim 7\,\rm d$.  The estimated
point-to-point photometric precision of the observations made by both
cameras, derived from the high-frequency noise, is $\sim 1.5\,\rm ppt$
(see also Section~\ref{sec:gran} and Fig.~\ref{fig:granspec}). Note
that we also created a combined dataset by concatenation of the
residuals from the two cameras, resulting in a set with a combined
length of 1051.4 days, and fill of 39.5\,per cent.

The residual lightcurves show evidence of long-period variability, of
amplitude several ppt, on timescales of a few hundred days.  The plot
demonstrates that observed trends appear to continue across
transitions in coverage from one camera to another, suggesting they
may well have a stellar component to them. It is interesting to note
that the observed variations are similar, both in amplitude and
timescale, to the long-period variations observed in Doppler velocity
by Irwin et al. (1989), Hatzes \& Cochran (1993) and Gray \& Brown
(2006). That said, there may be some instrumental contributions to the
long-period trends. A detailed survey on the long-period stability of
the \emph{SMEI} photometric data is currently the subject of ongoing
work (Spreckley \& Stevens, in preparation).


\begin{figure*}

\centerline{\epsfxsize=7.25cm\epsfbox{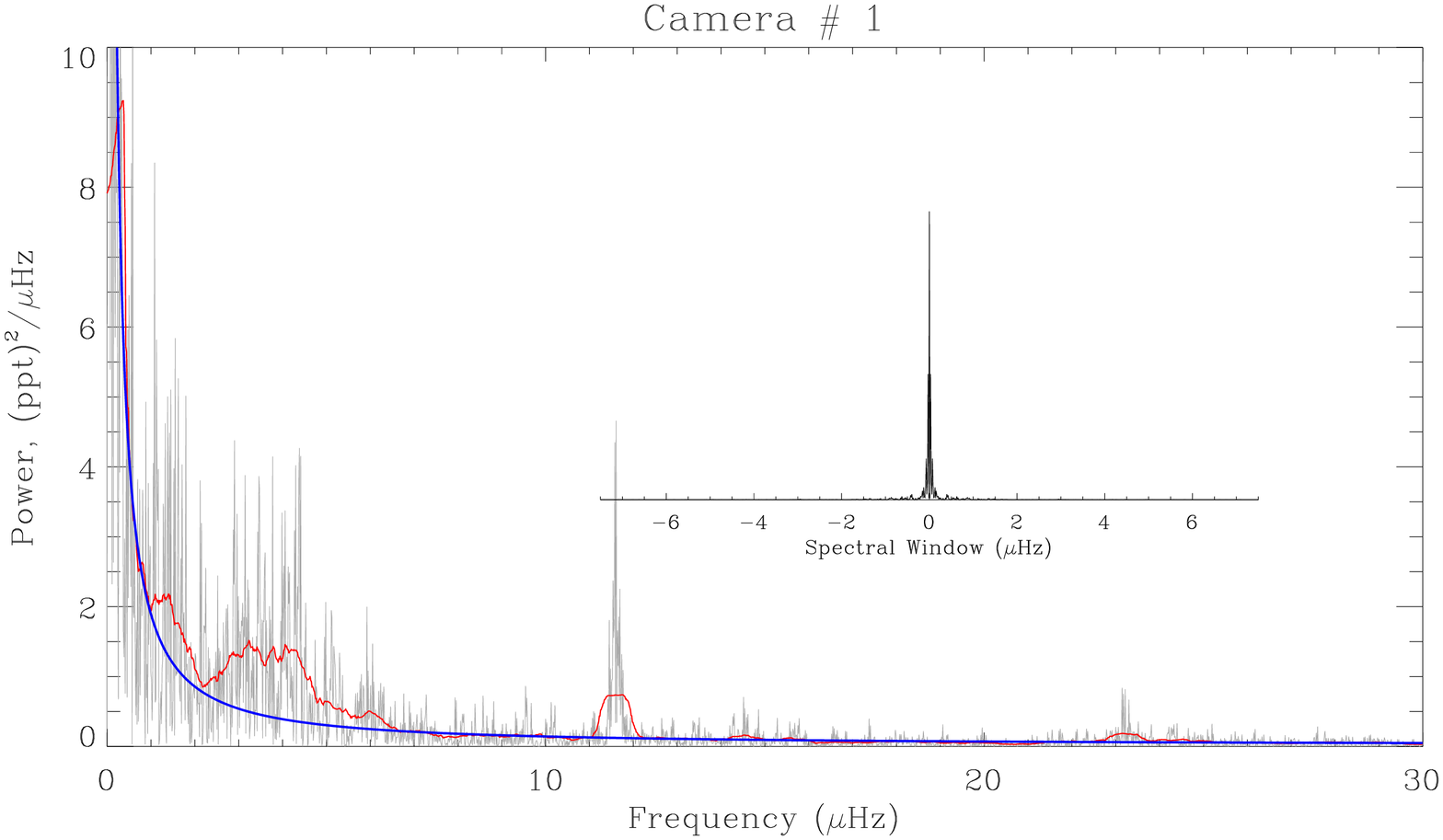}
            \epsfxsize=7.25cm\epsfbox{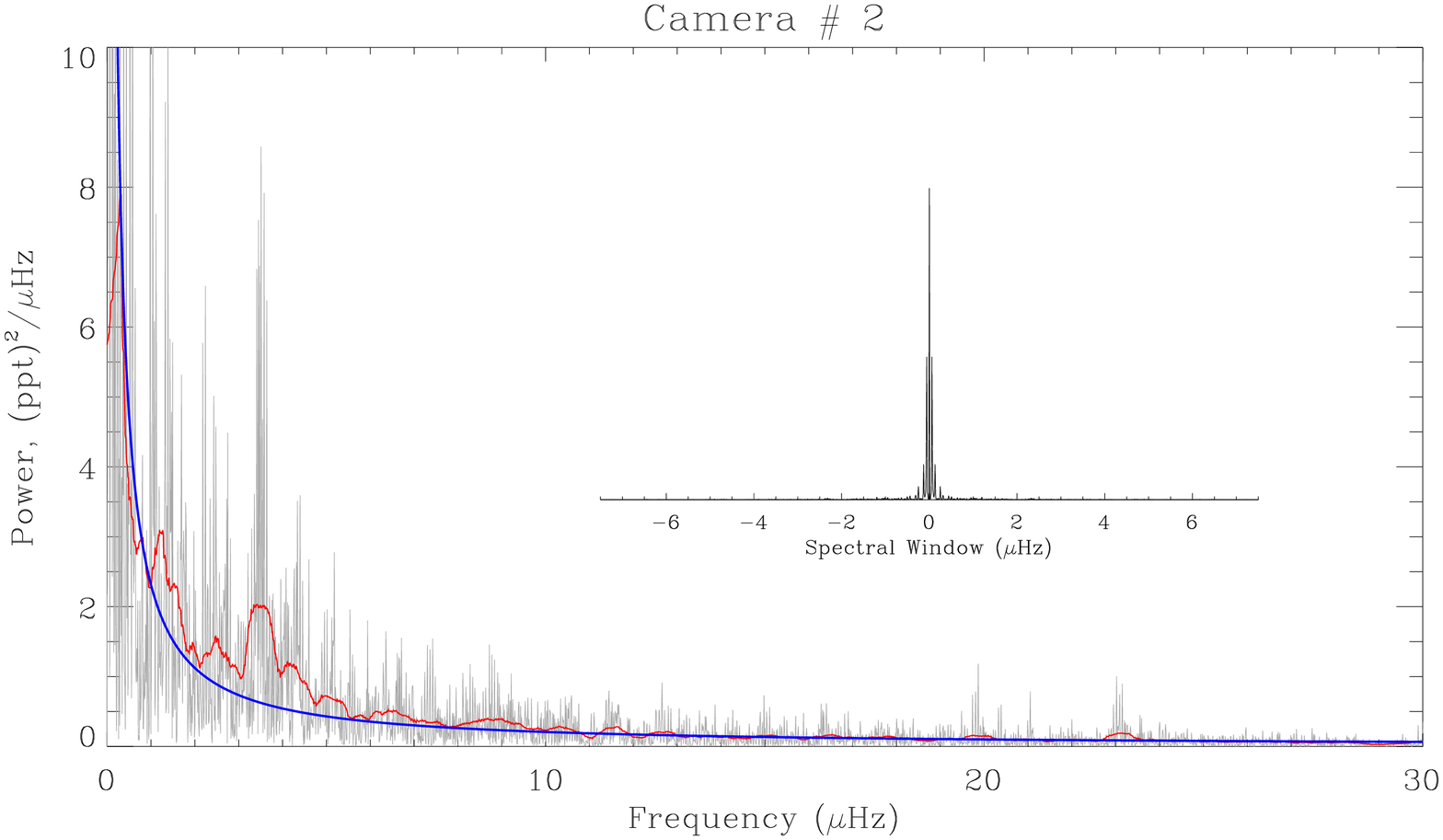}}
 
\caption{Power density spectra of the Camera \#1 (left-hand panel) and
  Camera \#2 (right-hand panel) residual lightcurves. The spectra are
  rendered in grey. Smoothed power spectra are shown as red lines. The
  blue lines show power law fits to the background (see
  Section~\ref{sec:anal}). The insets show spectral windows for each
  dataset, at the same horizontal frequency scale as the main plots.}

\label{fig:powspec} 
\end{figure*}


Power density spectra of the Camera \#1 and \#2 residual lightcurves
are shown in Fig.~\ref{fig:powspec} between 0 to $30\,\rm \mu Hz$.
The raw spectra are rendered in grey. The red lines show the results
of smoothing each spectrum with a boxcar filter of width $\sim
0.7\,\rm \mu Hz$.  The blue lines show power law fits to the
background (see Section~\ref{sec:anal}). The insets to both panels
show power spectra of the window functions. (The window spectra are
dominated by the effects of the 360 and 180-day gaps between the start
times of data segments in each lightcurve, giving peaks at separations
of 0.03 and $0.06\,\rm \mu Hz$, respectively.)

The power density spectra of both residual lightcurves show
concentrations of power, centered on about $4\,\rm \mu Hz$, which
appear to lie above the slowly varying background. A prominent peak
located at $\approx 3.5\,\rm \mu Hz$ in the Camera \#2 spectrum is
particularly striking (S/N$\approx$8 in power).  The locations in
frequency of these peaks are consistent with predictions for p modes
based on simple scaling relations that work well for Main Sequence,
subgiant and giant stars.  Bedding \& Kjeldsen (2003) have shown that
the frequency at which Sun-like oscillations have their maximum
observed amplitudes scales with the acoustic cut-off frequency, i.e.,
as $\sim M R^{-2}\,T^{-1/2}_{\rm eff}$. The radius and effective
temperature of Arcturus are both well determined, e.g., $R = 25.4 \pm
0.3\,\rm R_{\odot}$ (Gray \& Brown 2006) and $T_{\rm eff} = 4290 \pm
30\,\rm K$ (Griffin \& Lynas-Grey 1999). The mass is less well
constrained. Taking $M = 0.8 \pm 0.3\,\rm M_{\odot}$ (Bonnell \& Bell
1993), and scaling against the frequency of maximum power for radial
solar p modes of $3100\,\rm \mu Hz$, the prediction is that the
maximum of the Arcturus p-mode spectrum should lie at $\sim 4.5\pm
1.7\,\rm \mu Hz$.

The power in the peaks, in particular that in the $\approx 3.5$-$\rm
\mu Hz$ peak, is also consistent with the excess power observed by
Retter et al. (2003) in photometric observations of Arcturus lasting
19\,d, which were made by the \emph{Wide Field Infrared Explorer}
(\emph{WIRE}) satellite. Retter et al. attributed this excess power as
being due to either several p modes, or a single, more heavily damped
p mode. Because the \emph{SMEI} lightcurves are much longer than the
\emph{WIRE} lightcurve, giving superior resolution in frequency, we
are in a better position to determine the nature of the excess
power. We discuss the characteristics of the \emph{SMEI} peaks, and
possible interpretations, below.

Before that, we note that the Camera \#1 spectrum has a very prominent
peak at the diurnal frequency of $11.57\,\rm \mu Hz$. Moreover,
prominent harmonics of the diurnal frequency are present at other
frequencies in the Camera \#1 spectrum (e.g., the second harmonic at
$\sim 23.1\,\rm \mu Hz$). A few overtones of the diurnal frequency are
also present in the Camera \#2 spectrum, though typically at far less
significant levels. The peaks in both spectra come from weak
signatures of stray earthlight in the data (there are much weaker
diurnal signatures in the Camera \#1 and \#2 window functions). There
also appears to be a second lump of power between $1-2\,\rm \mu
Hz$. However, because of the sharply increasing background over this
region these features are not statistically significant (the
statistical tests are described below in Section~\ref{sec:anal}).

\section{Analysis of the power density spectra}
\label{sec:anal}

\begin{figure}

\centerline{\epsfxsize=7.25cm\epsfbox{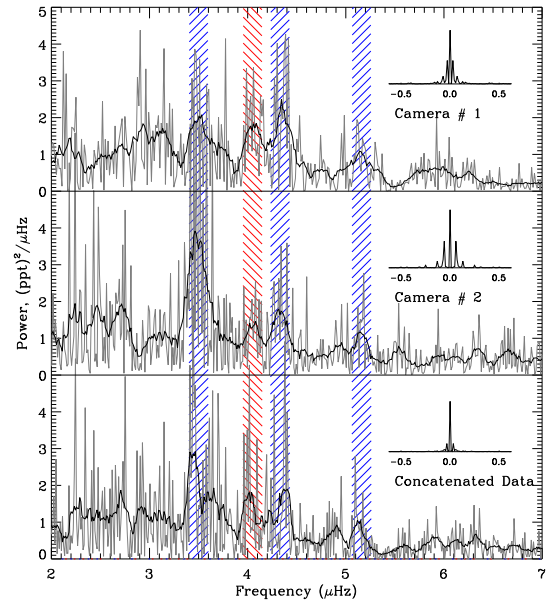}}
 
\caption{Power density spectra of the Camera \#1 (upper plot) and
Camera \#2 (middle plot) and concatenated (lower plot) residual
lightcurves, between 2 and $7\,\rm \mu Hz$. The insets show spectral
windows for each dataset.}

\label{fig:powboth} 
\end{figure}

Fig.~\ref{fig:powboth} shows the region of interest in power density
spectra of the Camera \#1, Camera \#2 and concatenated residual
lightcurves, between 2 and $7\,\rm \mu Hz$. The raw spectra are again
rendered in grey. The dark solid lines show the results of smoothing
each spectrum with a boxcar filter of width $\sim 0.14\,\rm \mu Hz$.

The hashed regions in Fig.~\ref{fig:powboth} have been added to guide
the eye to locations where peaks appear to be present at the same
frequencies in the spectra. Peaks marked by the forward-slanted
blue hashes are located at $\sim 3.5$, $\sim 4.35$ and $\sim 5.15\,\rm \mu
Hz$ and show an approximate equidistant spacing in frequency (see
Section~\ref{sec:modes} below). Both spectra also show a peak at $\sim
4.05\,\rm \mu Hz$ (backward-slanted red hashes) which does not respect the
near equidistant spacing between the other peaks. Each spectrum also
shows some other peaks in the range 2 to $3.2\,\rm \mu Hz$ which do
not coincide in frequency.


It it certainly tempting to identify the peaks marked by the
forward-slanted blue hashes as constituting a spectrum of overtones of
radial p modes. But is each of the peaks significant, i.e., could the
appearance of the spectra be replicated merely by the random
fluctuations of a non-resonant, broad-band background? We tested the
likelihood that either prominent spikes -- a spike being a prominent
power spectral density, or height, in a \emph{single} frequency bin --
or several prominent spikes lying in close proximity in frequency --
i.e., peaks, which could be the signature of a damped mode -- were
part of the slowly rising (in frequency), smooth background noise.
Our tests returned estimates of the probability of finding spikes or
peaks by chance anywhere in the spectrum (from zero frequency up to
the putative Nyquist frequency of $\sim 82.5\,\rm \mu Hz$). Low values
of the returned probabilities would flag spikes or peaks as being
deserving of further consideration as possible candidate modes. This
is the so-called null, or H0, hypothesis. We refer the reader to
Chaplin et al. (2002) and Appourchaux (2004) for full details on the
spike and peak tests.

Our tests demanded that we pre-whiten the spectra first to allow us to
determine heights of peaks relative to the local background levels,
which of course change with frequency. We obtained estimates of the
underlying backgrounds in two different ways, both of which involved
fitting models to the observed power density spectra: (i) by fits of a
simple power law $P(\nu) = a \nu^b + c$, where $a$, $b$ and $c$ were
coefficients to be fitted; and (ii) by fits of a three-component
model, including a component to represent explicitly granulation,
which is described more fully in Section~\ref{sec:gran} below. The
smooth background estimates given by fitting model (i) to each
spectrum are shown as blue lines in each panel of
Fig.~\ref{fig:powspec}.

In order to give robust estimates of the likelihood of spikes and
peaks occurring by chance we used Monte Carlo simulations of
artificial data, modulated by the window functions, to fix the
likelihood levels. In this way we were able to take proper account of
the influence of windows upon results. Full details on the Monte Carlo
simulations will be given in an upcoming paper. Here, we note that we
created one thousand independent artificial realizations of the
lightcurves for the two different models of the background described
in the previous paragraph, each of which contained \emph{no}
artificial p modes.  These data gave power density spectra closely
resembling the observed spectra.  By searching each of the artificial
spectra for prominent spikes and peaks, using the methods applied to
the real spectra, we were able to estimate likelihood distributions
for our tests, and to therefore assign significance levels to each of
the peaks in the real spectra.

We chose to flag as potential candidate modes those peaks which, when
tested, had estimated likelihoods of $\le 1$\,per cent of appearing by
chance anywhere in the spectrum. In demanding that tests return such
low likelihoods we sought to guard against possible false
detections. (See Broomhall et al. 2007 for more discussion on fixing
the threshold.) The peak at $\approx 3.5\,\rm \mu Hz$ passed the
significance threshold in the Camera \#2 and concatenated datasets.
We note that the peaks at $\sim 3.5$, $\sim 4.05$ and $\sim 4.35\,\rm
\mu Hz$ did reach the $\approx 1$\,per cent level in certain tests in
all spectra, but the likelihoods were notably less significant in
other tests (with results depending on how the data were
pre-whitened).

In summary, we conclude it is extremely unlikely that the Camera \#2
peak at $3.5\,\rm \mu Hz$ is due to noise. We feel confident in
identifying this peak as being the signature of a mode. The other
peaks in the range 2 to $7\,\rm \mu Hz$ are certainly worthy of note,
and may be the signatures of other modes; however, we cannot rule out,
with the same degree of confidence, that they are not simply part of
the background noise. That peaks appear at similar frequencies in the
Camera \#1 and \#2 spectra is intriguing. It is of course possible to
test the likelihood that peaks appear by chance in the same bin, or
bins, of either spectrum (Broomhall et al. 2007) (having first
truncated the Camera \#2 lightcurve to be the same length as the
Camera \#1 lightcurve). Since the data in the lightcurves are almost
completely independent in time, it is a trivial matter to find the
joint likelihood of occurrence. When we tested against the joint
likelihoods, the conclusions drawn from the independent tests were
upheld.

\section{Discussion: peaks as modes}
\label{sec:modes}


\begin{figure}

\centerline{\epsfxsize=7.0cm\epsfbox{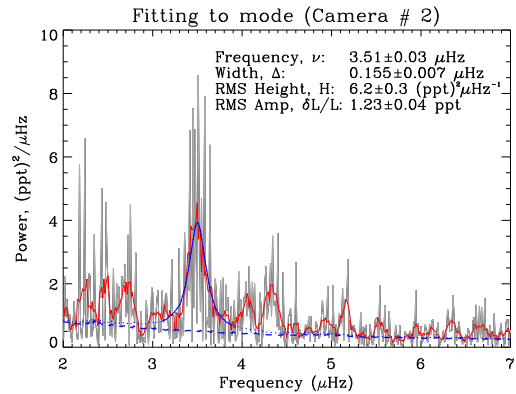}}
 
\caption{Fit to the mode seen in the Camera \#2 power density
spectrum. The raw spectrum is shown in grey and a 5-bin
boxcar-smoothed spectrum in red. The best-fitting model is plotted in
blue (with the solid part showing the range over which the mode was
fitted); the background is plotted as the thick dashed line.}

\label{fig:mode} 
\end{figure}


Our results in Section~\ref{sec:anal} above suggest it is likely that
the prominent peak located in the Camera \#2 spectrum at $3.5\,\rm \mu
Hz$ is the signature of a mode.  What are the basic parameters of this
mode? We have fitted the spectrum in the immediate vicinity of the
mode to a multi-component model comprised of a resonant profile, to
describe its peak, and a multi-component background. The obvious
choice for the resonant profile is a Lorentzian function. However, if
the $Q$-factor of the oscillation peak (i.e., frequency over
linewidth, $\nu/\Delta$) is low, the full classical resonant function,
which has terms in $\nu^4$, must be used. Here, the $Q$ factor is a
fairly modest $\approx\,25$, and we tried both functions. The fitting
model was also convolved with the Fourier transform of the window
function to allow for redistribution of power by the window (see
Fletcher 2007 for an in-depth discussion of the technique). Fits with
either resonant profile gave very similar best-fitting parameters for
the mode. We show the best fit to the peak, which incorporated the
full classical function, in Fig.~\ref{fig:mode}. The best-fitting mode
parameters are also shown on the figure.

The results on this mode allow us to make several important
remarks. The excellent frequency resolution of the data has allowed us
to make a direct measurement of the width, $\Delta$, of the mode,
whose value implies a lifetime $\tau = 1/(\pi \Delta)$ of $24 \pm
1\,\rm d$. The presence of significant width, even after allowing for
aliasing of power by the window function, implies the mode is stable,
and excited stochastically by convection. The $Q$-factor of the mode
peak is $23 \pm 1$. This is slightly lower than the $Q$ of
$\approx\,50$ found for modes on the red giant $\xi$\,Hya (Stello et
al. 2006); and similar to the $Q$ of modes on the red giant
$\epsilon$\,Oph (Barban et al. 2007) [$Q$ of $\sim 30$, determined to
a fractional precision of about 25\,per cent].

Assuming the $3.5\,\rm \mu Hz$ peak to be the strongest mode in the
spectrum, the observed RMS amplitude is found to be in approximate
agreement with the predicted amplitude given by the scaling relation
of Kjeldsen \& Bedding (1995). This relation uses the maximum
amplitude of radial solar p modes as a baseline calibration, and gives
amplitudes that scale linearly with the luminosity to mass ratio,
$L/M$. There is also a correction for the colour of the star (i.e.,
$T_{\rm eff}$) and the wavelength, $\lambda$, at which the
observations are made.  Taking the values given in
Section~\ref{sec:data} for $T_{\rm eff}$, $M$ and $R$, together with
$\lambda = 700\,\rm nm$, this being at the peak in sensitivity of the
\emph{SMEI} data, we obtain a predicted maximum RMS amplitude for
Arcturus of $(dL/L) = 1.1 \pm 0.5\,\rm ppt$.  This prediction agrees
within errors with our observed amplitude. Samadi et al. (2007) have
recently discussed how theoretical predictions for the amplitudes of
Sun-like p modes are affected by the description of the dynamic
properties of the turbulent convection. In one scenario, which Samadi
et al. favour, the amplitudes are found to scale like
$(L/M)^{0.7}$. Such a scaling would instead imply $(dL/L) = 0.2 \pm
0.1\,\rm ppt$ for Arcturus. This prediction differs significantly from
our observed RMS amplitude.

In order to provide further support for the conclusions given above,
we also conducted simulations like those outlined in
Section~\ref{sec:anal} but now with an artificial p mode included in
each artificial dataset, having the same parameters as our best fit
shown in Fig.~\ref{fig:mode}. We were thereby able to verify that
changes to the prominence and appearance of the 3.5-$\rm \mu Hz$ peak
in the Camera \#1, Camera \#2 and concatenated residual lightcurve
spectra, \emph{and} the spectra of individual segments from each
camera, were consistent with our inferred values for the mode
lifetime, amplitude and S/N.

What of the other peaks in Fig.~\ref{fig:powboth}? Those peaks marked
by the forward-slanted blue hashes show an equidistant spacing in
frequency, of just over $\sim 0.8\,\rm \mu Hz$. Taking the values
given in Section~\ref{sec:data} for $M$ and $R$, and scaling against
the mean observed spacing for radial solar p modes (of $135\,\rm \mu
Hz$), we get a predicted overtone spacing for Arcturus of $\sim
0.9\,\rm \mu Hz$. So, one possible explanation for these peaks is that
they correspond to three overtones of the fundamental radial
mode. However, we remind the reader that we feel we cannot rule out,
at least at a confidence level $\le 1$\,per cent, that the peaks are
not due to noise. Finally, we note the peak at $\sim 4.05\,\rm \mu Hz$
marked by the backward-slanted red hashes in both spectra in
Fig.~\ref{fig:powboth}. Could it be the signature of a non-radial
mode? Again, we advise caution in over interpreting these results, but
hope that improvements to the lightcurve pipeline may give cleaner
data, and a more definitive answer regarding the origins of this peak,
and the other peaks discussed above.

\section{Estimation of granulation parameters}
\label{sec:gran}


\begin{figure}

\centerline{\epsfxsize=6.5cm\epsfbox{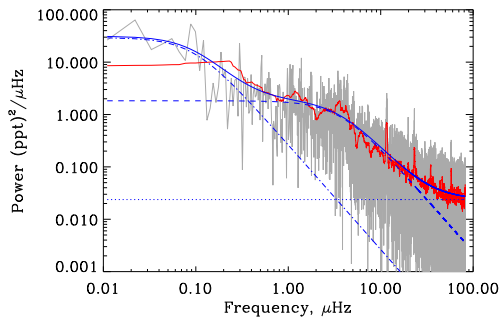} }
 
\caption{Power density spectrum of the concatenated residual
lightcurve (smoothed in red), showing three-component model fit to the
background.  The overall best-fitting curve is plotted as a dark blue
line, with contributions of individual components plotted as follows:
photon shot noise (dotted lines); granulation (dashed lines);
long-period noise, including stellar active-region noise, plus
instrumental noise (dot-dashed line).}

\label{fig:granspec} 
\end{figure}


Fig.~\ref{fig:granspec} shows the power density spectrum of the
concatenated residual lightcurve, with power now plotted on a
logarithmic scale. The smooth dark blue line is the best-fitting curve
of a three-component model used to describe the background power. The
model contains a flat (white) component to describe the photon shot
noise; and two power-law components to represent the contributions of
granulation and stellar active region noise.  It is important to
remember that some of the long-period variability in the lightcurves
(Fig.~\ref{fig:res}) may have an instrumental origin, and, as such,
the third component must also take account of instrumental
contributions. The second and third components were each described by
the power-law model of Harvey (1985) (see also Bruntt et
al. 2005). The Harvey model is specified by two parameters: a standard
deviation, $\sigma$, and a characteristic timescale, $\tau$. The
best-fitting power contributions of the individual components are
plotted as dotted (shot noise), dashed (granulation noise) and
dot-dashed (long-period noise) lines in Fig.~\ref{fig:granspec}.  The
unprecedented frequency resolution allows well-constrained estimates
of the granulation parameters to be extracted. The characteristic
`knee' in the power spectral density, which is given by the flattening
of the granulation power at low frequencies, is clearly apparent in
the spectrum (see region near $\approx 1\,\rm \mu Hz$). The Camera \#1
and \#2 spectra show the same feature, and are of similar appearance
to Fig.~\ref{fig:granspec}. The best-fitting granulation parameters
for the spectrum of the concatenated residual lightcurve are: $\sigma
= 4.56 \pm 0.06\,\rm ppt$ and $\tau=0.502 \pm 0.004\,\rm
days$. Comparison of the $\sigma$ and $\tau$ values from similar fits
to each of the Camera \#1 and \#2 spectra gives a guide to the actual
external precision in the parameters: $\approx$10\,per cent in
$\sigma$, and $\approx$10\,per cent in $\tau$.

\section{Conclusion}
\label{sec:conc}

From our analysis of photometric observations made by \emph{SMEI}, we
have identified a prominent mode of oscillation on Arcturus (see
parameters listed in Fig.~\ref{fig:mode}) and, by fitting its resonant
peak, have obtained a precise estimate of its damping time ($\tau = 24
\pm 1\,\rm d$). The excellent resolution in frequency is unprecedented
for near-continuous, high-cadence photometric studies of red giant
variability, and this is the first time an individual mode of
oscillation of Arcturus has been identified unambiguously.  Analysis
also hints at the possible presence of several radial-mode overtones,
and maybe some non-radial modes, at frequencies in the range $\approx
2$ to $5.5\,\rm \mu Hz$. We have also measured the granulation
properties of the star.

\section*{Acknowledgments}

NJT and SAS acknowledge the support of STFC. SAS also acknowledges the
support of the School of Physics \& Astronomy, University of
Birmingham. \emph{SMEI} was designed and constructed by a team of
scientists and engineers from the US Air Force Research Laboratory,
the University of California at San Diego, Boston College, Boston
University, and the University of Birmingham. We single out
A. Buffington, C. J. Eyles and S. J. Tappin for particular thanks.  We
also thank the referee for helpful comments.


\begin{thebibliography}{}

\bibitem[]{appourchaux04} Appourchaux T., 2004, A\&A, 428, 1039

\bibitem[]{barban07} Barban C., et al., 2007, A\&A, 468, 1033

\bibitem[]{bedding03} Bedding T. R., Kjeldsen H., 2003, PASA, 20, 203

\bibitem[]{bedding06} Bedding T. R., Kjeldsen H., 2006, in:
  SOHO18/GONG 2006/HELAS I, `Beyond the spherical Sun', August 2006,
  Sheffield, ed. M. Thompson, ESA SP-624, Noordwijk, Netherlands,
  p.~25.1

\bibitem[]{belmonte90} Belmonte J. A., Jones A. R., Pall\'e P.,
  Roca-Cort\'es T., 1990, ApJ, 358, 595
  
\bibitem[]{bonnell93} Bonnell, J. T., Bell, R. A., 1993, MNRAS, 264, 334

\bibitem[]{broomhall07} Broomhall A.-M., Chaplin W. J., Elsworth Y.,
  Appourchaux T., 2007, MNRAS, 379, 2

\bibitem[]{bruntt05} Bruntt H., Kjeldsen H., Buzasi D. L., Bedding
T. R., 2005, ApJ, 633, 440

\bibitem[]{chaplin02} Chaplin W. J., et al., 2002, MNRAS, 336, 979

\bibitem[]{dalsgaard04} Christensen-Dalsgaard J., 2004, Sol. Phys., 220, 137

\bibitem[]{deridder06} de~Ridder J., et al., 2006, A\&A, 448, 689

\bibitem[]{fletcher07} Fletcher S. T., 2007, PhD thesis, University of
Birmingham, UK

\bibitem[]{frandsen02} Frandsen S., et al., 2002, A\&A, 394, 5

\bibitem[]{gray06} Gray D. F., Brown K. I. T., 2006, PASP, 118, 1112

\bibitem[]{griffin99} Griffin R. E. M., Lynas-Grey A. E., 1999, AJ,
  117, 2998

\bibitem[]{harvey85} Harvey J., 1985, in: Future missions in solar,
  heliospheric and space plasma physics, eds. E. Rolfe, B. Battrick,
  ESA SP-235, Noordwijk, Netherlands, p.~199

\bibitem[]{hatzes93} Hatzes A. P., Cochran W. D., 1993, ApJ, 413, 339

\bibitem[]{hatzes94} Hatzes A. P., Cochran W. D., 1994, ApJ, 422, 366

\bibitem[]{hekker06} Hekker S., Aerts C., de~Ridder J., Carrier F.,
  2006, A\&A, 458, 931

\bibitem[]{innis88} Innis J. L., et al., 1998, in: Seismology of the
  Sun and Sun-like Stars, eds. V. Domingo, E. J. Rolfe, ESA SP-286,
  Noordwijk, Netherlands, p. 569

\bibitem[]{irwin89} Irwin A. W., Campbell B., Morbey C. L., Walker
G. A. H., Yang S., 1989, PASP, 101, 147

\bibitem[]{kjeldsen95} Kjeldsen~H., Bedding T. R., 1995, A\&A, 293, 87

\bibitem[]{merline95} Merline W. J., 1995, PhD thesis, Univ. Arizona

\bibitem[]{retter03} Retter A., Bedding T. R., Buzasi D. L., Kjeldsen
  H., Kiss L. L., 2003, ApJ, 591, L151

\bibitem[]{samadi07} Samadi R., Georgobiani D., Trampedach R., Goupil
  M. J., Stein R. F., Nordlund A., 2007, A\&A, 463, 297

\bibitem[]{stello06} Stello D., Kjeldsen H., Bedding T. R., Buzasi D.,
  2006, A\&A, 448, 709

\bibitem[]{stetson87} Stetson P. B., 1987, PASP, 99, 191

\end{thebibliography}
  \end{document}